\documentclass[conference]{IEEEtran}
\IEEEoverridecommandlockouts

\usepackage{cite}
\usepackage{amsmath,amssymb,amsfonts}
\usepackage{algorithmic}
\usepackage{graphicx}
\usepackage{textcomp}
\usepackage{xcolor}

\def\BibTeX{{\rm B\kern-.05em{\sc i\kern-.025em b}\kern-.08em
    T\kern-.1667em\lower.7ex\hbox{E}\kern-.125emX}}

\def\baselinestretch{0.991}
\begin{document}

\title{Direct-Conflict Resolution in Intent-Driven Autonomous Networks\\
\thanks{Author Idris Cinmere is funded for his Ph.D. by the Ministry of National Education in Turkiye.}
}
\author{\IEEEauthorblockN{1\textsuperscript{st} Idris Cinemre}
\IEEEauthorblockA{\textit{Dept. of Engineering} \\
\textit{King’s College London}\\
London, UK \\
idris.1.cinemre@kcl.ac.uk}
\and
\IEEEauthorblockN{2\textsuperscript{nd} Kashif Mehmood}
\IEEEauthorblockA{\textit{Dept. of Information Security and Communication Technology} \\
\textit{Norwegian University of Science and Technology (NTNU)}\\
Trondheim, Norway \\
kashif.mehmood@ntnu.no}
\and
\IEEEauthorblockN{3\textsuperscript{rd} Katina Kralevska}
\IEEEauthorblockA{\textit{Dept. of Information Security and Communication Technology} \\
\textit{Norwegian University of Science and Technology (NTNU)}\\
Trondheim, Norway \\
katinak@ntnu.no}
\and
\IEEEauthorblockN{4\textsuperscript{th} Toktam Mahmoodi}
\IEEEauthorblockA{\textit{Dept. of Engineering} \\
\textit{King’s College London}\\
London, UK \\
toktam.mahmoodi@kcl.ac.uk}

}

\maketitle

\begin{abstract}
As network systems evolve, there is an escalating demand for automated tools to facilitate efficient management and configuration. This paper explores conflict resolution in Intent-Based Network (IBN) management, an innovative approach that holds promise for effective network administration, especially within radio access domain. Nevertheless, when multiple intents are in operation concurrently, conflicts may emerge, presenting a significant issue that remains under-addressed in the current literature. In response to this challenge, our research expands the range of conflict resolution strategies beyond the established Nash Bargaining Solution (NBS), to incorporate the Weighted Nash Bargaining Solution (WNBS), the Kalai-Smorodinsky Bargaining Solution (KSBS), and the Shannon Entropy Bargaining Solution (SEBS). These methods are employed with the objective to identify optimal parameter values, aiming to ensure fairness in conflict resolution. Through simulations, it is demonstrated that distinct antenna tilt values are yielded as the respective solutions for each method. Ultimately, based on Jain’s Fairness Index, the KSBS is identified as the most equitable method under the given conditions.
\end{abstract}

\begin{IEEEkeywords}
Intent-based networking (IBN), Cooperative Bargaining Problem, Nash Bargaining Solution (NBS)
\end{IEEEkeywords}

\section{Introduction}

Intent-based networking (IBN) has emerged as a promising enabler of autonomous network management \cite{rfc9315, 9925251}. It focuses on desired outcomes rather than specific configurations; thus, representing a paradigm shift towards more flexible, agile, and simplified network management \cite{MEHMOOD2023109477}. Given the inefficiencies and high costs associated with the manual management of wireless networks, especially in Radio Access Networks (RAN), a transition towards automated network management is essential.  The concept of Self-Organizing Networks (SON), endorsed by the wireless operator consortium NGMN \cite{alliance2007ngmn} and the 3GPP \cite{network2011self}, has been proposed as a solution to significantly reduce operational costs. Traditionally, the management of radio parameters has been conducted through these rule-based SON systems. With the introduction of artificial intelligence (AI), AI-based approaches have replaced rule-based SONs with more advanced cognitive autonomous networks (CAN) \cite{canbook}. CANs leverage machine learning algorithms to optimize network performance, efficiently allocate resources, and respond adaptively to dynamic network conditions by changing radio parameters.

Building on the evolving landscape of network management, the implementation of intent-driven autonomous networks promises to simplify even further the RAN management for Mobile Network Operators (MNOs) \cite{8968429}. By employing high-level abstractions of intent, network administrators can focus on what needs to be achieved, rather than the complexities of how to achieve it \cite{rfc9315}. This approach not only enhances the effectiveness of network management but also alleviates the need for low-level device configurations and implementation details, and reduces potential errors or misconfigurations. 


In the context of network management utilizing IBN, situations may arise where multiple intents are received simultaneously, leading to potential conflicts~\cite{conflict-etsi}. A notable form of such a conflict is referred to as a direct target conflict \cite{mwanje2022intent}. This reveals when multiple intents, in their pursuit of achieving distinct outcomes, impose contradictory requirements on a specific network control parameter. Such a scenario manifests when one intent may necessitate a reduction in the value of a particular network control parameter, whilst a concurrent intent demands an increase of the same parameter's value.

Baktir \textit{et al.}\cite{9844074} address conflicts in intent-based systems characterized by multiple closed-loop control processes. In this framework, each intent initiates closed-loop management to achieve its predefined Key Performance Indicators (KPIs). Conflicts arise when the actions of one closed loop negatively impact another loop's ability to meet its KPIs. To resolve these conflicts, the proposed method employs fitness values or penalties associated with each intent, representing the cost of failing to fulfill or partially fulfilling their KPIs. An intent manager computes an overall penalty based on these individual penalties. During conflict resolution, an evaluation agent assesses the impact of potential KPI changes, selects actions aimed at minimizing the predicted overall penalty, and actuation agents execute these actions within the controlled environment.


Moreover, Perepu \textit{et al.}\cite{10001426} simulate a scenario where an Intent Manager must fulfill three distinct intents for different services, with packet priority and maximum bit rate as the controlling parameters. Conflicts emerge when optimizing one target KPI negatively affects another. To address this, the authors employ a model-free multi-agent reinforcement learning approach, where each agent is responsible for tuning a parameter related to the intent-defined objective based on intent priorities (expressed as penalties). This approach leads to a proportional degradation of lower-priority intents to prioritize higher-priority ones.


The study in \cite{9829768} focuses on a closed-loop control-based system where each KPI is managed by a cognitive function (CF). These CFs learn the dependencies between the managed KPI and Network Control Parameters (NCPs) by observing the effects of NCP adjustments on KPI outputs. The study assumes the existence of a single intent that activates two distinct functions, Mobility Robustness Optimization (MRO) and Mobility Load Balancing (MLB), which control the same parameters in a conflicting manner within a cognitive autonomous network. Conflicts arise when the target parameter values of these functions do not align. To address this contradiction problem, the authors propose the use of the Nash social welfare function as a solution to determine the optimal value of the parameter. The estimated target values derived from the functions and the suggested optimum value are then transmitted to the MNO for decision-making. As the number of contradictions increases, the complexity of managing this decision-making process may escalate, potentially leading to performance reductions due to the absence of an automated decision process.

Network operators are often presented with scenarios where they have to accommodate contrasting demands, necessitating a careful balancing act of their resources in order to simultaneously fulfill multiple intents. This challenge, although identified, has not been comprehensively addressed in the literature.
The primary tool utilized for conflict resolution in this context has been the Nash Bargaining Solution (NBS). Its application has marked a significant step forward in the literature, providing a valuable foundation for addressing these complex problems. This work extends on the existing bargaining solutions for conflict resolution in IBN as follows:

\begin{itemize}
    \item To broaden our investigation to encompass a broad range of bargaining solutions, including the Weighted Nash Bargaining Solution (WNBS), the Kalai-Smorodinsky Bargaining Solution (KSBS), and the Shannon Entropy Bargaining Solution (SEBS).
    \item To identify optimal parameter values using these methods with the objective of ensuring Jain-based fairness in intent conflict resolution framework.
    \item To design a validation model for the RAN cellular network environment in order to verify the conflict resolution performance of proposed methodologies.
\end{itemize}



This paper is organized as follows: Section II presents the cooperative bargaining problem and the bargaining solution concepts. Section III introduces the system model, including a detailed description of the simulation scenario and the intent-based bargaining problem. Section IV evaluates performances based on simulations. Finally, Section V concludes the paper and highlights future research directions.


\section{The Cooperative Bargaining Problem}

The cooperative bargaining problem, $\mathcal{B}$, a fundamental concept in cooperative game theory and economics, involves two or more parties negotiating to reach a mutually beneficial agreement on an alternative or accepting their differing viewpoints by selecting a disagreement point. The problem is given as 
\begin{equation}
\mathcal{B} =  \left( S,\mathbf{d} \right)
\end{equation}
\noindent where $S$ is the nonempty convex and closed set of all possible agreements, called the feasible utility set and $\mathbf{d}$ is the disagreement point, which is the outcome that the players will receive if they cannot reach an agreement. Each element in $S$ represents an agreement the players can potentially reach and the disagreement point is the reference point from which players evaluate the benefits of a possible agreement.

Given a set of players $N = \{1, 2, ..., n\}$ and a disagreement point $\mathbf{d} \in \mathbb{R}^n$, where $\mathbf{d} = [d_1, d_2, ..., d_n]$ represents the utility for each player in the case of disagreement. Define a feasible set $S \subset \mathbb{R}^n$, representing all the possible outcomes of the bargaining, where for all $\mathbf{x} \in S$, $x_i \geq d_i$ for $i = 1, 2, ..., n$.

\subsection{The Nash Bargaining Solution (NBS)}
NBS aims to find an agreement $\mathbf{x}^{\text{NBS}} \in S$ that maximizes the product of the players' utilities, taking into account their utilities at the disagreement point. The utilities of the players $\left( i=1,2, ..., n \right)$ are represented by $x_i$ where $x_i \in \mathbf{x}$ and the disagreement point, $\mathbf{d}$, is an array of the minimum agreeable utilities of players ( i.e. $\mathbf{d}= \left[ x^{min}_1, x^{min}_2, \ldots, x^{min}_n \right]$). The NBS can be formally defined as follows \cite{ebrahimkhani2021bargaining}:
\begin{equation}
\mathbf{x}^{\text{NBS}} = \arg\max_{\mathbf{x} \in S}  \prod_{i=1}^{n} \left(x_i- d_i \right)
\end{equation}
where the solution, $\mathbf{x}^{\text{NBS}}$, is chosen such that it maximizes the product of the utilities' gains (compared to the disagreement point) for all players. 
\begin{equation}
\mathbf{x}^{\text{NBS}} = [{x}^{\text{NBS}}_1, {x}^{\text{NBS}}_2, ..., {x}^{\text{NBS}}_n]
\end{equation}
where ${x}^{\text{NBS}}_i = u_i({p}^{\text{NBS}})$. The utility of the $i$-th player, $x_i$, is defined by $u_i(p)$ at a given control parameter value $p$.  The NBS is a solution concept that helps to find a stable and fair outcome in such problems by maximizing the product of players' utilities gains over the disagreement point. 
WNBS is an extension of the NBS that takes into account the relative importance (priority) of each player in the bargaining process \cite{9051725}. By incorporating weights, $w_i$, we can reflect the difference in bargaining power between the players $\left( i=1,2, ..., n \right)$. 
\begin{equation}
\mathbf{x}^{\text{WNBS}} = \arg\max_{\mathbf{x} \in S}  \prod_{i=1}^{n} \left(x_i- d_i \right)^{w_i}
\end{equation}
\noindent where $\sum_{i=1}^n w_i=1$ and  $w_i \geq 0$. 
The WNBS aims to find an agreement $\mathbf{x}^{\text{WNBS}} \in S$ that maximizes the weighted product of the players' utilities, taking into account their utilities at the disagreement point. By adjusting the weights, $w_i$, we can control the balance between the players' utilities in the bargaining solution, reflecting their relative bargaining power (i.e. their relative priorities).

\subsection{Kalai-Smorodinsky Bargaining Solution (KSBS)}
KSBS is a solution concept in cooperative game theory that generalizes the Nash bargaining solution. It ensures that the bargaining outcome is proportional to the difference between each player's maximum and minimum payoffs, which are represented by their utility functions and disagreement points, respectively. The goal of KSBS is to distribute gains fairly among the players while maintaining proportionality to their disagreement points, thus promoting equitable outcomes.

For an $n$-player bargaining scenario, the KSBS proportionality conditions can be expressed as \cite{borgstrom2007rate}:
\begin{equation}
    \frac{x_1^{\text{KSBS}} - d_1}{x_1^* - d_1} = \cdots = \frac{x_n^{\text{KSBS}} - d_n}{x_n^* - d_n}
\end{equation}
where $x_1^*,\ldots, x_n^*$ are the outcomes that maximize the respective utility functions for each player. $\mathbf{x}^{\text{KSBS}}$ is the solution that satisfies the KSBS. The proportionality conditions ensure that the relative gains for each player in the bargaining solution are proportional to their maximum achievable gains beyond their disagreement points. This results in a fair distribution of benefits among the players while respecting their individual preferences and negotiation power.

\subsection{Shannon Entropy Bargaining Solution (SEBS)}

In the context of bargaining problems, an alternative approach to finding a fair utility allocation is by maximizing the Shannon entropy, also known as the Theil index, which quantifies the diversity or randomness in the allocation. The Theil index is a measure of inequality that is based on concepts from information theory, particularly entropy. It is often used in the context of income distribution, but it can also be applied to other types of allocation problems, such as utility allocation in bargaining games. In this context, the Theil index can be used to find a fair allocation of utility among the involved parties
and it can be defined as the optimization problem as follows \cite{sampat2019fairness}:
\begin{align*}
\mathbf{x}^{\text{SEBS}} &= \arg\max_{\mathbf{x} \in S} h_S \\
\text{subject to} \quad & \sum_{i \in N} (x_i - d_i) = c
\end{align*}
\noindent where $c$ is a constant value, $h_S$ is the Shannon entropy (Theil index) of the utility allocation, given by
\begin{equation}
h_S = -\sum_{i \in N} p_i \log p_i,
\end{equation}
and $p_i$ is the proportion of utility gain for player $i$, given by
\begin{equation}
p_i = \frac{(x_i - d_i)}{\sum_{i \in N}(x_i - d_i)}
\end{equation}
where $\sum_{i \in N}p_i =1$, $N$ represents the set of players and utility of player $i$ is depicted by $x_i$. Also, the disagreement point for player $i$, which is the minimum utility, is denoted as $d_i$. $\mathbf{x}^{\text{SEBS}}$ is the solution that satisfies the SEBS.

\section{System Model}
We consider a system model with two intents which are in a direct target conflict. After a significant earthquake with a high magnitude, the mobile network of a city is facing significant challenges. A considerable portion of base stations have been damaged, leading to a reduction in the availability of the network. Additionally, there has been a significant surge of voice calls and text messages, which has further strained the network's capacity.

It is assumed that two different intents are received; 
\begin{itemize}
  \item Intent A:\emph{ Reestablish connectivity to areas where residents have lost coverage due to network outages},
  \item Intent B:\emph{ Enhance the network capacity in the region that encompasses the coordination center}. 
\end{itemize}
\noindent The Intent A seeks to restore connectivity to areas where a considerable number of residents have lost coverage due to network outages or congestion. The focus is on ensuring that emergency communications, such as calls to first responders or texts between family members, are reliable and efficient. Simultaneously, Intent B aims to serve users seeking access to video streaming services to stay informed about the earthquake's impact and receive updates on relief efforts. In regions where there is a significant demand for streaming services, such as disaster response coordination centers, network operators may need to optimize capacity and ensure reliable video streaming.

\begin{figure}[t]
	\centering
	\includegraphics[width=3.5in]{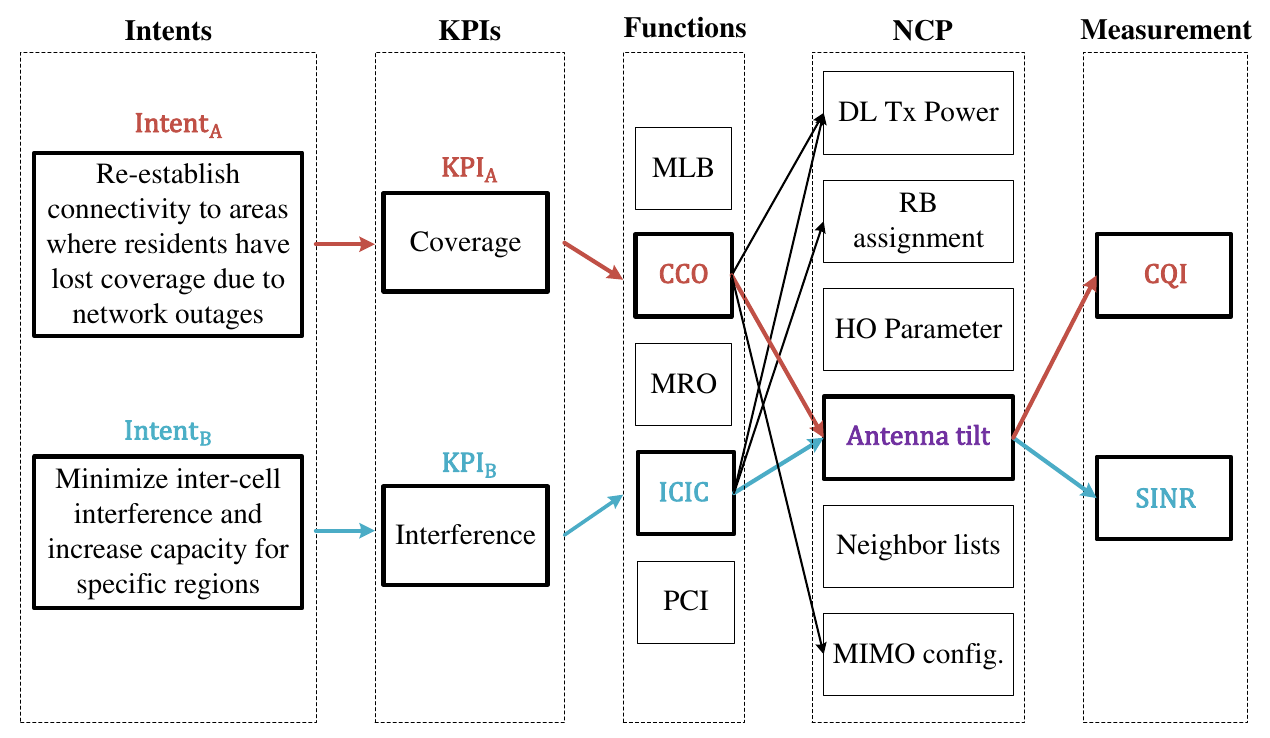}
	\caption{Relations between Intents, Functions, Network Control Parameters.}
	\label{f:p1}
\end{figure}

To address these two intents, coverage and interference KPIs must be modified. The Coverage and Capacity Optimization (CCO) and Inter-Cell Interference Coordination (ICIC) functions could be employed to manage these KPIs. The CCO function can increase transmission power levels to reduce outage probability at the cell edge, while the ICIC function can decrease power transmission levels to minimize interference and maximize capacity. Alternatively, the CCO function can adjust the antenna tilt by increasing tilt to expand the coverage area, enhancing coverage in affected regions. Simultaneously, the ICIC function can decrease the tilt to concentrate the signal in a smaller area and reduce inter-cell interference. The relationship between intents, KPIs, SON functions (i.e MLB), and multiple tunable NCPs, such as transmission power (Tx power) and antenna tilt is given in Fig. \ref{f:p1}. These elements have a complex, interconnected relationship where adjusting a group of NCPs can influence a function value, and conversely, a single NCP change may impact several functions \cite{5990492}. 

This scenario exemplifies a \emph{direct target conflict} which refers to a situation where two or more objectives have contradictory goals concerning a specific network parameter \cite{mwanje2022intent}. In this instance, one intent seeks to decrease the value of the specific network control parameter, while another intent necessitates an increase in the same parameter. The conflicting intents in this scenario center around common parameters such as downlink transmission power and antenna tilt adjustments. With limited network resources and infrastructure, it becomes challenging to simultaneously satisfy both demands. 

\begin{figure}[b]
	\centering
	\includegraphics[width=3.5in]{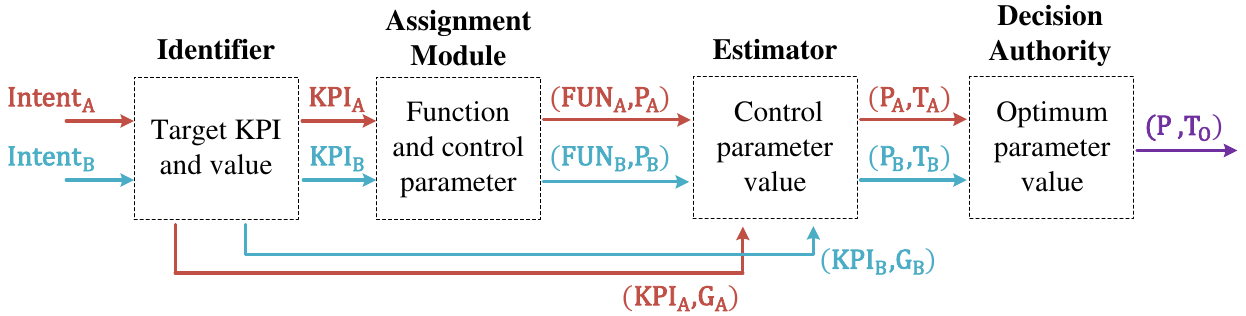}
	\caption{Conflict Resolution Workflow.}
	\label{f:pr}
\end{figure}

Fig. \ref{f:pr}  provides a detailed representation of the workflow associated with the proposed conflict resolution method.  Upon receiving the intents(Intent A, Intent B), the Identifier proceeds to delineate the KPIs ($KPI_A, KPI_B$) and their corresponding targeted values ($G_A, G_B$) encapsulated from the intents. Following this process, the Assignment Module conveys the functions ($FUN_A, FUN_B$) employed, as well as the network control parameter ($P_A, P_B$) that requires tuning, both of which are determined by the target KPIs. The Estimator holds the duty of discerning the precise value for each control parameter ($T_A, T_B$). This determination is contingent upon factors such as the function type employed, the specific KPI in question, and the designated target value for the KPI. Lastly, the Decision Authority bears the responsibility of executing the bargaining solutions and identifying equitable options within the pool of all existing methods. The aim here is to ensure an unbiased selection process, as the Decision Authority seeks to decide the optimum value of the control parameter ($T_O$) that maximizes fairness in the application of the various methods.

\subsection{Intent-based Bargaining Problem}

For the defined scenario, we consider a situation where two conflicting intents arise: restoring connectivity and ensuring efficient emergency communications (Intent A), and minimizing inter-cell interference and optimizing capacity for video streaming services (Intent B). To find an optimal parameter value that balances these conflicting intents, the bargaining solution concepts defined in the previous section, such as the NBS, KSBS, and SEBS, are employed. 
Let $x_a$ represent the utility for Intent A, and $x_b$ represent the utility for Intent B and define target control parameter values for each intent as $T_a$ and $T_b$. In a direct conflict scenario, the optimum value should be  $T_a \leq T_o \leq T_b$. 

\subsubsection{The Nash Bargaining Solution (NBS)} In \cite{9829768}, NBS is proposed for contradiction management in intent-driven cognitive autonomous RAN without considering disagreement points. 
The goal is to find an optimal parameter value that balances these conflicting intents and maximizes the product of the utilities. To apply the NBS concept, we first need to define the feasible utility set $S$ and the disagreement point $\mathbf{d}$;
\begin{itemize}
  \item {\emph{Feasible utility set, $S$:}
In this case, the feasible utility set $S$ consists of all possible pairs of utilities $(x_a = u_a(T), x_b = u_b(T))$ for the range of antenna tilt adjustments $T$.}
  \item {\emph{Disagreement point, $\mathbf{d}$:}
The disagreement point represents the minimum agreeable utilities for both intents. It can be denoted as $\mathbf{d} = [x_a^{min}, x_b^{min}]$. }
\end{itemize}

Now we can apply the NBS for a two-player bargaining game to find the optimal tilt value, $T^{\text{NBS}}$, as
\begin{equation}
 \arg\max_{T} \left[ (u_a(T) - u_a^{min}(T)) \times (u_b(T) - u_b^{min}(T)) \right].
\end{equation}
In addition, the WNBS concept for a two-player bargaining game with the weights $w_a$ and $w_b$ can be utilized as
\begin{equation}
 \arg\max_{T} \left[ (u_a(T) - u_a^{min}(T))^{w_a} \times (u_b(T) - u_b^{min}(T))^{w_b} \right]
\end{equation}
\noindent to find the optimal tilt value, $T^{\text{WNBS}}$, where the weights $w_a$ and $w_b$ represent the relative importance or bargaining power of Intent A and Intent B, respectively, such that $w_a, w_b \geq 0$ and $w_a + w_b = 1$.
The objective is the identification of the optimal antenna tilt adjustment that maximizes the NBS function while considering the relative importance of the intents. For instance, if the priority is to emphasize emergency communications (Intent A), a higher weight can be assigned to $w_a$.

\subsubsection{The Kalai-Smorodinsky Bargaining Solution (KSBS)} 

The KSBS proportionality conditions for a two-player bargaining game:
\begin{equation}
\frac{u_a(T^{\text{KSBS}}) - u_a^{min}(T)}{u_a^*(T) - u_a^{min}(T)} = \frac{u_b(T^{\text{KSBS}}) - u_b^{min}(T)}{u_b^*(T) - u_b^{min}(T)} 
\end{equation}
\noindent where the maximum achievable utility for Intent A, $u_a^*(T)$, is the maximum value of $u_a(T)$ in the range of antenna tilt adjustments, and the maximum achievable utility for Intent B, $u_b^*(T)$, is the maximum value of $u_b(T)$ in the same range. The optimal $T^{\text{KSBS}}$ value will balance the conflicting intents while maintaining proportionality to the disagreement points and fairly distributing gains among the intents.

\subsubsection{The Shannon Entropy Bargaining Solution (SEBS)} 
To apply the SEBS concept, we first need to define the utility proportions for each intent and the entropy function;
\begin{itemize}
  \item {\emph{Utility proportions, $p_a$ and $p_b$:} Using the definition of $p_i$, the utility proportions for Intent A and Intent B are defined as:
\begin{equation}
p_a = \frac{(u_a(T) - u_a^{min}(T))}{(u_a(T) - u_a^{min}(T)) + (u_b(T) - u_b^{min}(T))}
\end{equation}
and
\begin{equation}
p_b = \frac{(u_b(T) - u_b^{min}(T))}{(u_a(T) - u_a^{min}(T)) + (u_b(T) - u_b^{min}(T))}
\end{equation}}
  \item {\emph{Shannon entropy function, $h_S$:} Using the utility proportions for Intent A and Intent B, the Shannon entropy function is defined as:
\begin{equation}
h_S = -[p_a \log p_a + p_b \log p_b]
\end{equation}}
\end{itemize}
Now, we can solve the following optimization problem:
\begin{align}
T^{\text{SEBS}} &= \arg\max_{T} \, -[p_a \log p_a+ p_b \log p_b] \\
\text{s.t.} \quad & (u_a(T) - u_a^{min}(T)) + (u_b(T) - u_b^{min}(T)) = c.
\end{align}
The optimal $T^{\text{SEBS}}$ value will balance the conflicting intents while ensuring a fair allocation of utility among the intents based on the proportion of utility gains.
\subsection{Jain-based Optimum Value} 
The Jain's Fairness Index (JFI) is often used to measure the fairness of resource allocation in a system. The formal definition of JFI, $J$, is defined as
\begin{equation}
J(\mathbf{x}) = \frac{(\sum_{i=1}^{n} x_i)^2}{n \sum_{i=1}^{n} x_i^2}
\end{equation}
\noindent where $n$ users with allocation $x_i$ for each $i \in \{1, 2, ..., n\}$. This index evaluates the equality of resource distribution. The closer the index is to 1, the more equally the resources are distributed. Conversely, the closer the index is to $\frac{1}{n}$, the more unevenly the resources are distributed. The optimal value of tilt is determined based on maximizing the JF among the solutions defined previously. 

\section{Performance Evaluation}

\subsection{Simulation setup}

\begin{figure}[t]
	\centering
	\includegraphics[width=3.5in]{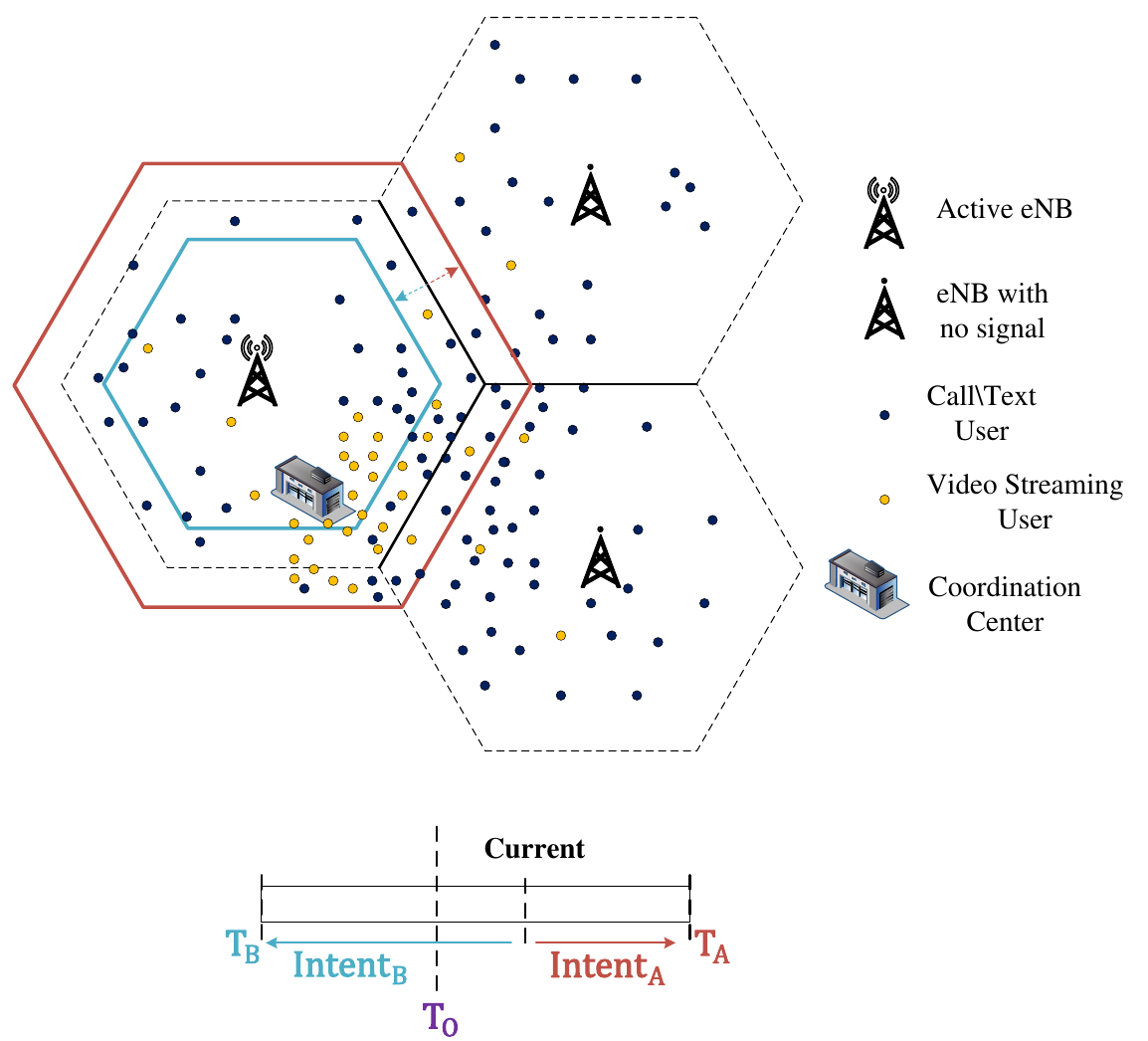}
	\caption{Simulation Scenario.}
	\label{f:Sce}
\end{figure}

A disaster recovery cellular network model is considered for modeling conflicts with an IBN approach. In the simulation scenario as shown in Fig. \ref{f:Sce}, three eNB sites are assumed, two of which are inactive due to earthquakes. Differentiation is service expectations is achieved by assigning users in two groups with call-text and video streaming services. Out of $100$ users, $80$ are call-text users, while the remaining $20$ engage in video streaming. This simulation scenario is evaluated using the ns-3 LENA platform, according to the parameters specified in Table 1.  

The system receives two concurrent intents for tuning the network control parameter, antenna tilt, to cater to the needs of video streaming users and call-text users. The transmission power of the antenna is maintained at a constant $46$ dBm, while the mechanical tilt of the antenna is varied within a range of $0^\circ$ to $15^\circ$, with incremental adjustments of $1^\circ$ at each step.
\begin{table}[b]
\caption{Simulation Parameters}
\label{table:simulation_parameters}
\centering
\begin{tabular}{l l}
\hline
Parameter                & Value                             \\ \hline
No. eNB sites            & 3                               \\
Sectors per site         & 3                               \\
No of CallText UEs                  & 80                               \\
No. of Video UEs                  & 20                               \\
Tx Power                 & 46 dBm                            \\
Path loss model          & 3GPPPropagationLossModel        \\
Mobility model for CallText           &  ConstantPosition                     \\
Mobility model for Video          & SteadyStateRandomWaypointMobilityModel                    \\
Scheduler                & Proportional fair                  \\
Shadow Fading            & Log-normal, std=8dB               \\
AMC model                & PiroEW2010                      \\
Cell layout radius       & 500m                              \\
Bandwidth                & 5MHz                              \\
No. of RBs               & 25; RBs per RBG:2                 \\
Horizontal angle $\phi$  & $-180^\circ \leq \phi \leq 180^\circ$ \\
Half power beamwidth     & Vertical 10$^\circ$: Horizontal 70$^\circ$ \\
Antenna gain $G_0$       & 10 dBi                            \\
Vertical angle $\theta$  & $-90^\circ \leq \phi \leq 90^\circ$ \\
Side lobe level          & Vertical -18 dB : Horizontal -20 dB \\
Side lobe level$_0$      & -30 dB                             \\
Actions (tilt)           & 0$^\circ$ -- 15$^\circ$: Granularity 1$^\circ$ \\
Simulation time          & 10s                               \\ \hline
\end{tabular}
\end{table}

\subsection{Simulation results}

The Signal-to-Interference-plus-Noise Ratio (SINR) values are computed for the video streaming users located near the coordination center. Concurrently, the Channel Quality Indicator (CQI) values are calculated for call-text users, who were primarily situated near the edge and outside of the coverage area. Note that the video streaming users are mobile, while the call-text users maintain a constant position.

\begin{figure}[t]
	\centering
	\includegraphics[width=2.8in]{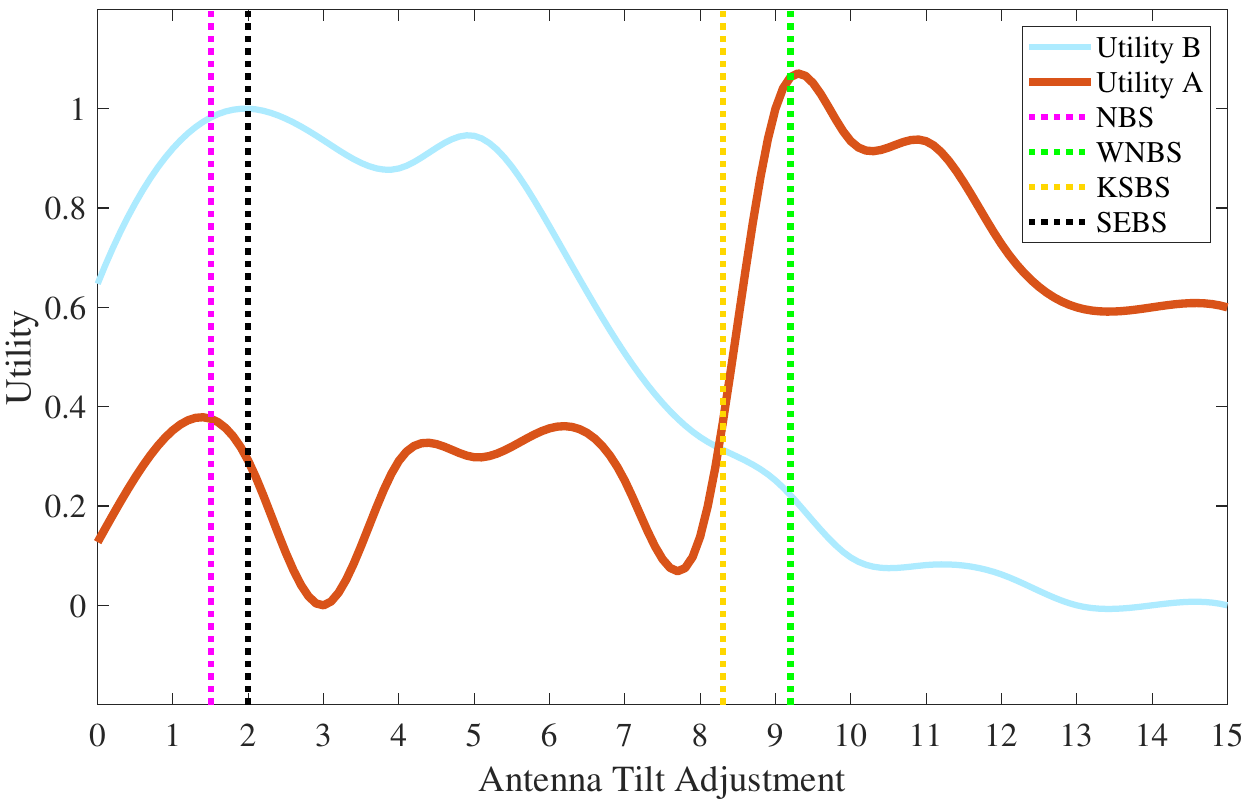}
	\caption{Variation of SINR (Intent B) and CQI (Intent A) Utility Functions with Antenna Tilt and Corresponding Bargaining Solutions.}
	\label{f:pro}
\end{figure}

Upon obtaining the simulation results, both SINR and CQI values are normalized to have the same unit, thus serving as a utility function. This normalization is performed based on
\begin{equation}
u_a = \frac{SINR - min(SINR)}{max(SINR)- min(SINR)} 
\end{equation}
and
\begin{equation}
u_b = \frac{CQI- min(CQI)}{max(CQI)- min(CQI)}. 
\end{equation}
Once the utility functions are established, the methods described as a solution in the previous section are applied. The resultant outcomes are presented in Fig. \ref{f:pro}. The utility of the CQI (i.e. Utility A) maintains consistency until the tilt angle reaches around $8^\circ$. Beyond this point, it noticeably increases as the antenna begins to engage targeted call-text users situated on and beyond the cell coverage boundary. On the other hand, the utility of the SINR (i.e. Utility B) remains at a higher value until the tilt is about $6^\circ$. Post this, there is a gradual decline due to the reduced signal power received by video streaming users closer to the base station, as they are affected by the increase in tilt values. Thus, it is appropriate to state that Intent A is oriented towards achieving higher tilt values, while Intent B focuses on attaining lower tilt values.

As depicted in Fig. \ref{f:pro}, each bargaining solution proposes a unique antenna tilt value when presented with two conflicting intents. Note that to demonstrate the impact of prioritization for the method WNBS, it is assumed that Intent A has been given a higher priority, denoted by a weight of $w_a=0.8$. Furthermore, when both priorities are set to the same value (i.e., $w_a = w_b = 0.5$), the WNBS produces the same result as the NBS.
The JFI is utilized to discern the optimal solution. Table II showcases the JFI values for the various methods, with the KSBS method emerging as the most equitable, as evidenced by its closest approximation to 1.
\begin{table}[h]
\caption{Fairness Results}
\label{table:simulation_parameters}
\centering
\begin{tabular}{l l l}
\hline
Methods        &Tilt value       & JFI Value                             \\ \hline
NBS             &$1.5^\circ$      & 0.8338                               \\
WNBS            &$9.2^\circ$      & 0.6997                              \\
KSBS            &$8.3^\circ$      & 0.9937                              \\
SEBS            &$2^\circ$      & 0.7682                             \\
 \hline
\end{tabular}

\end{table}

\section{Conclusion}

This paper is an exploration into the resolution of direct target conflicts in intent-driven autonomous networks. Alternative bargaining solutions, extending beyond the commonly used NBS, such as the WNBS, the KSBS, and the SEBS, are investigated. Through simulation, it is found that each method yields distinct antenna tilt values as their respective solutions. Based on JFI under the given conditions, the KSBS is identified as the most equitable method. These findings prompt further exploration in this field, particularly into the potential use of learning algorithms to optimize fairness by combining different bargaining methods.
\ifCLASSOPTIONcaptionsoff
\newpage
\fi
\bibliographystyle{IEEEtran}
\bibliography{IEEEabrv,reff}
\end{document}